\documentstyle[prl,aps]{revtex}
\input{epsf}
\begin{document}
\title{Evolution on a Smooth Landscape}
\author{David A. Kessler}
\address{Minerva Center for the Study of Mesoscopics, Fractals, and Neural
Networks and Dept. of Physics, Bar-Ilan Univ., Ramat Gan, Israel}
\author{Herbert Levine, Douglas Ridgway and Lev Tsimring}
\address{Institute for Nonlinear Science,
University of California, San Diego La Jolla, CA  92093-0402}
\maketitle
\begin{abstract}
We study in detail a recently proposed simple discrete model for evolution on 
smooth landscapes. An asymptotic solution of this model for long times is 
constructed. 
We find that the dynamics of the population are governed by
correlation functions that although being formally down by powers of $N$ 
(the population size) nonetheless control the evolution process after a very
short transient. The long-time behavior can be found analytically
since only one of these higher-order correlators (the two-point function)
is relevant. We compare and contrast the exact findings derived herein with
a previously proposed phenomenological treatment employing
mean field theory supplemented with a cutoff at small population density.
Finally, we relate our results to the recently studied case of mutation on 
a totally flat landscape.  

\ 

\end{abstract}
\section{Introduction}
In a recent paper, \cite{virus1} we introduced a model to describe evolution 
on a smooth landscape.  This model was motivated by the results 
of recent
experiments\cite{Holland95} on the evolution of fitness in populations of RNA viruses. 
In these experiments, the population shows secular changes in birth rate,
which is identified by the biologists as the logarithm of fitness.
These experiments showed a rapid initial increase of birth rate, followed by
a sharp transition to a regime of significantly slower, approximately linear, 
increase.  The initial rapid rise in birth rate was interpreted by us to be
a result of the exponentially quick dominance of the population by the most 
``fit'' (most rapidly reproducing) members of the initial population. These
most ``fit'' individuals produce many more offspring, so they quickly dominate,
in accord with a simple picture of the working of selection in a population.
The subsequent slow rise was interpreted as being due to the effect of 
mutations which, as the mutation rate is low, act on a slower time scale.

The relative smoothness of the increase of birth rate in this second, linear,
regime indicates that the ``fitness'' landscape must be fairly smooth.  It
is clear from many studies of evolution\cite{Eigen,Schuster,Kaufman} in 
rough, ``glassy'' 
landscapes, that
increases in fitness in such cases are expected to be sudden, 
with long intervals of little or no improvement in between. While it is
clear that the population should smoothly ``climb the hill'' in a smooth
landscape, no attention seems to have been focused on the precise dynamics
of such hill-climbing.  In particular, what is the speed with which the
population climbs, and on what does it depend?

To begin to address these questions, we formulated \cite{virus1} a very 
simple discrete model of evolution in a smooth landscape. This model
exhibited both the initial rapid ``collapse'' of the population onto
its most fit members, and the subsequent smooth rise in birth rate which
characterized the RNA virus experiments (see Fig 1.).  
In addition, we constructed a continuum mean-field treatment which, when 
suitably doctored with a cutoff, appeared to give a qualitatively
correct description of the dynamics.  As we were concerned with the gross
qualitative features of the model and its possible relevance to the 
experiment, we did not investigate the precise extent
of the apparent agreement nor indicate in any detail
as to where the cutoff came from. 

In this paper, we return to the
analysis of our discrete model.  Our analysis provides an exact description 
of the dynamics at asymptotically long times.  This asymptotic dynamics 
indicates that the population does indeed increase its birth rate in a
linear fashion, and provides a prediction of the rate of such increase.
It also provides a description of the stochastic departures from this
linear behavior. The most striking result of this analysis is that
the rate of increase of birth rate (the velocity of climbing of the
fitness hill) depends crucially on the population size (increasing
essentially linearly with the population). Naive mean-field theory is
however completely independent of $N$, the population size.
Indeed, we show explicitly how 
naive mean-field theory gives rise to a finite-time divergence (or in another
variant of the model, an exponential blowup) of the fitness and how
the true dynamics cures this unphysical behavior through terms which
are formally lower order in $1/N$.  These terms become relevant on
relatively short time-scales and drive the system to an $N$ dependent
asymptotic state. It is important to note that even the cutoff version of
mean-field theory, which does not suffer from the blowup phenomenon,
does not correctly predict this asymptotic state, due to the neglect of
a specific two-body correlation term which dominates at long-time. This 
diagnosis of the failings of mean-field
theory is the second major product of our analysis.  

\begin{figure}
\centerline{\epsfxsize = 3.in \epsffile{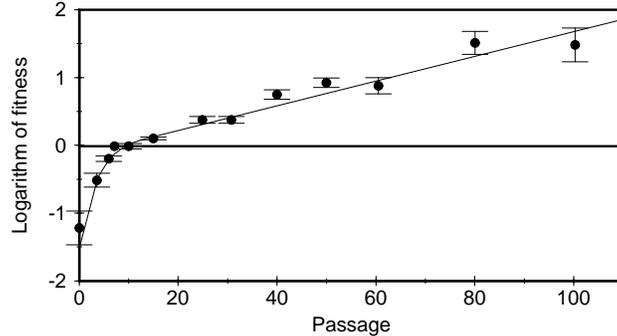}}
\caption{Evolution of fitness of MARM clone during the transmission
series of 80 passages on HeLA cells (Fig. 2b of \protect\cite{Holland95} 
with permission from PNAS)}
\label{fig1}
\end{figure}

\section{Preliminaries}
We begin our exposition by presenting the discrete model of evolution
which forms the basis of our discussions.  We also outline the naive
mean-field theory and the modified cutoff version we analyzed in our previous
work \cite{virus1}.

\subsection{The Discrete Model}
The discrete model whose behavior we will primarily focus on is a
slight variant of that discussed in our earlier work.  The original
model, as well as another variant, the flat landscape model,
can be easily analyzed in a manner parallel to that with which
we shall treat the version we present here.  The discussion of these other
variants we will thus postpone to a later section.

Our model consists of a fixed population of N individuals.  Although
in the experiments the population varied between two essentially
fixed values, the fixed population constraint is obviously the simpler
case to study first, and seems to preserve the qualitative features
of the more general case.  The individuals are characterized by their
birth rates, which the experimentalists associate with the logarithm of the
``fitness'' of the individual.  Each individual gives birth at his 
characteristic
rate in a standard Poisson process. The baby, due to mutation, may
have a birth rate which is $\pm \Delta$ that of his parent. The overall
mutation probability for a baby is a constant $\mu$. By rescaling time
by $\Delta$, we may take the change in the baby's birth rate to be $\pm 1$,
so that the birth rates are restricted to be integers. We do not allow
a baby to mutate to $0$ birth rate, so in fact the birth rates are all
natural numbers. As we work at fixed population size $N$, each birth
is occasioned by the unfortunate demise of some member of the group.  The
victim is chosen at random from among all members of the population 
(including the new-born baby) independent of their fitness rating.

To get some idea of the dynamics of this model, we show in Figure 2 the
results of a simulation of this model with a population size of $N=50$.
Starting from a population of identical individuals,
the average mean-birth-rate (note: average means over
different realizations, mean denotes the mean for all individuals in a
given simulation) accelerates quickly, followed by a gradual
deacceleration to a regime of constant velocity, i.e. a linear increase of
mean-birth-rate with time. Looking also at the
variability of the population, we calculate the average sample width-squared,
which increases rapidly from zero and then saturates at some finite value. 
We will see later that the transient phase is complicated,
but the asymptotic state is always one of constant velocity.

\subsection{Mean-Field Theory}
It is easy to construct a mean-field theory for the simple process outlined
above.  Let $P(n,t)$ be the number of individuals with birthrate $n$ at
time $t$.  Then, in the absence of mutations, it is clear that
\begin{equation}
\dot P = (n-\bar n) P(n)
\end{equation}
where $\bar n$ is the instantaneous mean value of $n$.  The presence of
mutations means that a fraction $\mu$ of the $nP(n)$ newborns will
leave birth-rate $n$, half moving up to $n+1$ and half down to $n-1$. Thus,
\begin{equation}
\label{mfdisc}
\dot P(n) = (n-\bar n) P(n) - n\mu P(n) + \frac{1}{2}\mu\left[(n+1)P(n+1) - 
(n-1)P(n-1)\right]
\end{equation}
or, in continuum language, replacing the discrete $n$ by $x$
\begin{equation}
\label{mfcont}
\dot P(x) = (x-\bar x) P(x) + \frac{1}{2}\mu (xP(x))''
\end{equation}
This equation is essentially identical to that of the mean-field theory for
Diffusion-Limited Aggregation (DLA)\cite{dla} in the far-field (large-$x$) 
regime. It is not surprising, therefore, that both the mean-field DLA 
\cite{tu} and our mean-field Equation (\ref{mfdisc}) share the property
that an initial pulse accelerates to infinite velocity in finite time.  
We will return to our
mean-field Equation (\ref{mfdisc}), and in particular demonstrate
explicitly the finite-time singularity, in Section VI.

\vspace{-1.75in}
\centerline{\epsfxsize = 6.in \epsffile{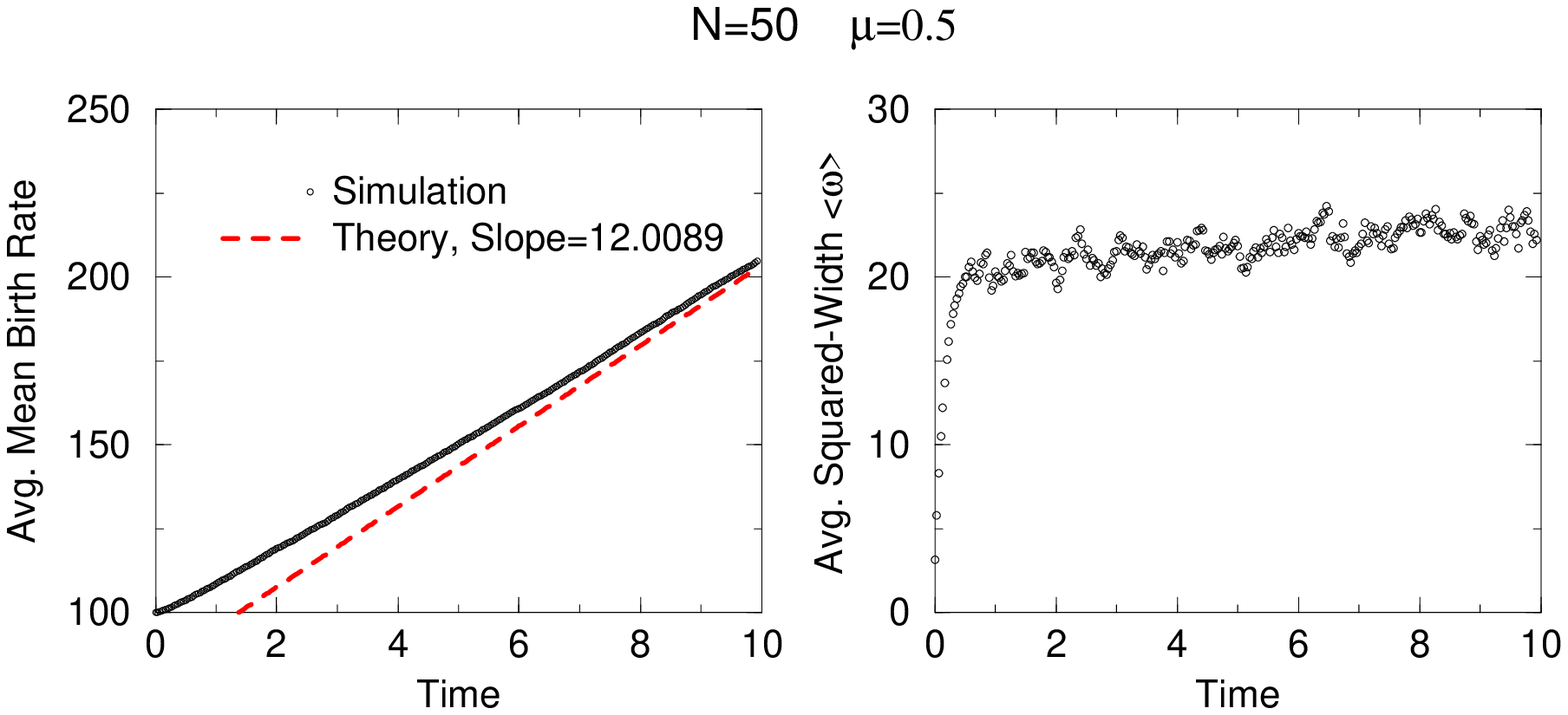}}
\begin{figure}
\caption{a) Mean birth-rate for population of $N=50$ individuals
as a function of time, with mutation rate $\mu=.5$, averaged over 200
realizations. Dotted line shows
theoretical predicted rate of increase of mean-birth-rate. \newline
b) Width-squared from same simulation.}
\end{figure}

For the moment, it is important to inquire as to the origin of this 
unphysical behavior. From a mathematical 
standpoint, the  essential problem is the term $xP$, which is unbounded.  
Physically, although
it is extremely unlikely for an individual to find itself far in front
of the rest of the population, should it happen, then that individual's growth
rate would be enormously greater than anyone else's.  Thus, this very
unlikely scenario makes a significant contribution to the ensemble average
$P(x)$ since its contribution is the product of the (tiny) probability of this 
configuration times the (huge) number of individuals at this site.  In the
model, however, the contribution of such an outlyer to the ensemble average 
remains negligible since the total number of individuals in any given 
configuration is limited to $N$.  From this discussion, however, we can
surmise that the velocity of propagation should be a strongly increasing
function of $N$.  Since mean-field theory is a theory for {\em infinite}
$N$, mean-field theory is in some sense giving the correct answer, which
is an infinite velocity at infinite $N$.  While perhaps correct in this
sense, such an answer is of course of little use.

There are various prescriptions one might employ to remove this ugly feature
of the mean-field theory, some of which have been suggested in the context
of mean-field DLA \cite{tu,cutoff}. In our earlier work on a variant of
this model
\cite{virus1}, we employed the stratagem of cutting off the $(x-\bar x)P$ term
when $P(x)$ was under some threshold value $P_{*}$. It is clear from the
outset, however, that for this system such a resulting theory cannot give 
rise to a
steadily-propagating solution, in contradiction with simulation.  Furthermore,
it is not clear how to justify this prescription (or any of the others that
have been suggested) on a first-principles basis. Thus, we leave mean-field
theory for the interim, and study in detail some simpler, though nontrivial 
limits of the model, those of a very small number of individuals
and the case of very small mutation rate $\mu$.

\section{Simple Limiting Cases}

In this section, we discuss the simplest possible cases for the model as
described above, that of populations of either one or two individuals or
alternatively the case of extremely small mutation rate. As
we shall see, this exercise allows us to determine exactly the nature of the
long-time state and is furthermore quite useful in pointing out what one needs
to do to make sense of the utter failure of simple mean field ideas.

\subsection{The Case of $N=1$}
As an introduction to the case of 2 individuals, let us quickly see
what happens for one individual.  Although trivial to work out, it
is actually a useful signpost on the road to the more interesting
case of $N=2$.

For $N=1$, we can write down the exact master equation for the Markov
process.  As there is only one individual, a configuration is labeled
by the position of that one particle, so we can express everything in
terms of $f(n,t)$, the probability that the individual is at site $n$
at time $t$. As the only way the individual can move to a new spot is
by giving birth, which occurs at a rate $n$ and then dying in favor of
his child ( a 50\% probability) so that his child can survive and, with 
probability $\mu$, step to
the left or right. Thus, the exact master equation reads
\begin{equation}
\dot f(n) = -\frac{1}{2}\mu n f(n) + \frac{1}{4}\mu\left[(n+1)f(n+1)
+ (n-1)f(n-1)\right]
\end{equation}
or, in continuum terms, 
\begin{equation}
\dot f(x) = \frac{1}{4}\mu (xf(x))'' = \frac{1}{4}\mu\left(2f' + xf''\right)
\end{equation}
This is very reminiscent of the mean-field equation, Eq. \ref{mfcont}
except for the extra factor of $1/2$ in the diffusion constant, but,
more importantly, the total absence of the troublesome $xP$ term! The reason
is clear: with only one individual, there can be nobody out in front of the
pack out-reproducing everyone.  In fact, the birth process itself is 
completely irrelevant
in changing the configuration, as one of the two individuals existing
post-birth is immediately tagged for death, leaving the configuration
unchanged.  

As we shall soon see, this absence of the $xP$ term is a very general result,
and, in fact, the equation of motion for long times is, in the general case,
exactly the same as for the $N=1$ case, up to changes in the values of the
coefficients.

\subsection{N=2: The Small $\mu$ Limit}
The case of $N=2$ is less trivial.  A general configuration is specified
by the positions of both individuals, so that it constitutes a 2-dimensional
field.  However, things simplify tremendously in the small $\mu$ limit.
The point is that if $\mu$ were exactly $0$, then the population would
quickly collapse, with both individuals finding themselves at the same
location (most likely that of the fitter of the two). For small $\mu$, every
so often one of the two begins to wander away, but it can't get far 
before the population collapses again.  So, for small $\mu$, we need
consider only 2 types of configurations:
1)the two individuals on the same site, and 2) the
individuals on nearest-neighbor sites.  Call the probability the individuals are
on the same site $x$, $f(x)$.  The probability that they are on $x$ and $x+1$
we call $g(x)$.
The master eqn. for the process is then:
\begin{eqnarray}
\dot f(x) &=& - \frac{4}{3}\mu x f(x) 
               + \frac{1}{3}(1-\mu)\Big(xg(x-1) +  xg(x)\Big)
               + \frac{\mu}{6}\Big((x+1)g(x) + (x-1)g(x-1)\Big)\nonumber \\
\dot g(x) &=& \frac{2}{3}\mu\Big(xf(x) + (x+1)f(x+1)\Big)
               - \frac{1}{3}(1-\mu)\Big((x+1)+ x\Big)g(x)
               - \frac{\mu}{6}\Big((x+1) + x\Big)g(x)
\end{eqnarray}
Notice we have forbidden the transitions from ``$g$'' to the state with
a space between the 2 individuals, to conserve probability. We can of course
combine terms above, but it is written so that each term in the $f$ equation
corresponds directly to the same term in the $g$ eqn.
As we have argued, $g$ is small, of order $\mu$, so since we are working 
only to linear
order in $\mu$, we shall drop all terms involving $\mu g$. Then the above
simplifies nicely to
\begin{eqnarray}
\dot f(x) &=& - \frac{4}{3}\mu x f(x) + \frac{1}{3}\Big(
                xg(x-1) +  xg(x)\Big) \nonumber \\
\dot g(x) &=& \frac{2}{3}\mu\Big(xf(x) + (x+1)f(x+1)\Big)
               - \frac{1}{3}\Big((x+1)g(x)+ xg(x)\Big)
\end{eqnarray}
Notice now that there is an exact time-independent and $x$-independent
solution to these 
eqns., namely $g=2\mu f = const$. Of course, this solution is not
normalizable, but as we will see, this is what the system is approaching
for long times. This motivates us to guess that
\begin{equation}
\label{slave}
g(x) \approx 2 \mu f(x) + \mu f'(x) \left(A_1 + \frac{A_2}{x}\right) + 
\mu f''(x) B
\end{equation}
or, equivalently (to the order we will need, (see below))
\begin{equation}
2\mu f(x) \approx g(x) + g'(x) \left(C_1 + \frac{C_2}{x}\right) + g''(x) D
\end{equation}
Notice that this is an expansion in derivatives, which are assumed
to decrease in magnitude with increasing order.  The justification for
this is the statement above, that $f$ and $g$ are approaching constants
at long time, so the expansion is essentially one in inverse powers of $t$.
We will see this explicitly later. It is also a large $x$ expansion.
In order to compute the velocity, we will need to keep terms up to
$x^{-1}$ in $f'$, $g'$ and up to $x^0$ in $f''$, $g''$. We can relate
the coefficients $A$ and $B$ to $C$ and $D$ by substituting the first 
expression
in the second and expanding.  We find $A_1=-2C_1$, $A_2=-2C_2$, and
$B=-2C_1^2-2D$. 

The idea now is to substitute the above relation in the equation of motion for 
$f$ and $g$. We note that $\dot f$ is proportional to $\mu$, but
$\dot g$ has terms zeroth order in $\mu$. For this to be consistent with
Eq. (\ref{slave}) relating $g$ to $f$,
we must make these zeroth order terms vanish.  This gives us equations for the
coefficients $C$ and $D$. In more detail,
\begin{eqnarray}
\dot g(x) &=& \frac{1}{3}\left[(2x+1)g'\left(C_1 + \frac{C_2}{x}\right) + g''D + 
             (x+1)(g' + C_1 g'') + \frac{x}{2}g''\right] \nonumber \\
\ &=& \frac{1}{3}\left[xg'(2C_1 + 1) + g'(C_1 + 2C_2 + 1) 
             + xg''(2D + C_1 + \frac{1}{2})\right]
\end{eqnarray}
Thus, $C_1=-\frac{1}{2}$, $C_2 = - \frac{1}{4}$, and $D=0$, so that
$A_1=1$, $A_2=\frac{1}{2}$, and $B=\frac{1}{2}$.  Substituting the 
relation for $g$ in term of $f$ in the eqn. for $\dot f$ yields 
\begin{eqnarray}
\dot f(x) &=& \frac{2\mu x}{3}\left[f'\left(A_1 + \frac{A_2}{x}\right) + f''B\right] -
             \frac{\mu x}{3}[2f' + A_1 f''] + \frac{\mu x}{6}2f''\nonumber\\
\ &=& \frac{\mu x}{3} \left[f'\left(2A_1 + \frac{2A_2}{x} - 2\right) + f''(2B - A_1 + 1)\right]\nonumber \\
\ &=& \frac{\mu}{3}[f' + xf'']
\end{eqnarray}
The last eqn. directly implies that the velocity $v$ is $\mu /3$, since
\begin{equation}
v = \frac{d}{dt}\int{dx x f(x)} = \frac{\mu}{3}\int{xf' + x^2f''} = \frac{\mu}{3}
\end{equation}
where the last result comes from integration by parts and the fact that $f$
is normalized to unity at lowest order.  Furthermore, there 
exists a similarity solution for this $f$ equation, (about which we will
discuss more later) with $x$ scaling like
$t$, so that a long-time expansion is also a large-$x$ expansion as advertised.
Also, as advertised, the form of the equation of motion for $f$ is exactly
as in the $N=1$ case.  What we have therefore, is essentially a bound-state
of the two individuals propagating together at constant velocity (on average)
with the center of mass performing (inhomogeneous) diffusion.  The 
squared-width of this
bound-state, the expectation value of the squared-distance between the two
individuals, is $g/(f+g)$, which to the order we are calculating is
simply 2$\mu$.

\subsection{General $N$: The Small $\mu$ Limit}
We can extend the calculation in the previous section to arbitrary $N$. To
lowest order in $\mu$, we have to consider only those states where the 
individuals are distributed between 2 nearest-neighbor sites.  We label the 
probability
of being in the state with
$N-k$ individuals at $x$ and $k$ individuals at $x\pm 1$ by 
$g^\pm_k(x)$, ($ 0 < k < N/2$). The probability of the state with all $N$ individuals at $x$ is 
$f(x)$. If $N$
is even, we also have the symmetric state $g_{N/2}(x)$, but for our purposes
it is enough to do the odd $N$ case. Again, we can write
down the master equation for the process, ignoring transitions out to other
states, which are down by additional powers of $\mu$. The equations of motion 
then read 
\begin{eqnarray}
\dot f(x)&=&\frac{1}{N+1}\bigg[-N^2\mu xf(x) + (N-1)x\Big(g_1^+(x) + g_1^-(x)\Big)\bigg]
\nonumber\\ &\ &\nonumber\\
\dot g_1^\pm(x) &=& \frac{1}{N+1}\bigg[\frac{1}{2}N^2\mu xf(x) 
           - (N-1)(2x\pm 1)g_1^\pm(x) + 2(N-2) x g_2^\pm(x)\bigg] \nonumber \\ 
&\ &\nonumber\\
\dot g_k^\pm(x) &=& \frac{1}{N+1}\bigg[-(N-k)k (2x\pm 1) g_k^\pm(x)
                 +(N-k-1)(n+1) x g_{k+1}^\pm(x)  \\
&\ &\ \ \ \ \ \ \ \ \ \ \ +(N-k+1)(n-1) (x \pm 1) g_{k-1}^\pm(x)\bigg]\nonumber\\
&\ &\nonumber\\
\dot g_{\frac{N-1}{2}}^\pm(x) &=& 
\frac{1}{N+1}\bigg[-\frac{N-1}{2}\cdot\frac{N+1}{2}(2x\pm 1) 
                                    g_{\frac{N-1}{2}}^\pm(x) 
+\frac{N-1}{2}\cdot\frac{N+1}{2} x g_{\frac{N-1}{2}}^\mp(x\pm 1)
 \nonumber\\
&\ &\ \ \ \ \ \ \ \ \ \ \ +\frac{N-3}{2}\cdot\frac{N+3}{2} (x\pm 1) g_{\frac{N-3}{2}}^\pm(x)\bigg ] \nonumber
\end{eqnarray}
where the middle equation holds for $k=2,\ldots,(N-3)/2$.
All the $g$'s are again 
slaved to $f$, so we write
\begin{equation}
g^{\pm}_k(x) =\mu f(x)\left(A_k^0  \pm \frac{A_k^1}{x} + \frac{A_k^2}{x^2} \right)
+ \mu f'(x)\left(\pm B_k^0 + \frac{B_k^1}{x}\right) + \mu f''(x)C_k
\end{equation}
We have used the reflection symmetry to determine the $\pm$ signs in this
equation.  Essentially, they came from the fact that the $\pm 1$'s in the
equations of motion are down by $1/x$ from the leading terms.
The procedure is exactly the same as before.  The coefficients $A_k^i$, 
$B_k^i$, and $C_k$ are determined by the requirement that the $\dot g_k^\pm$
must all vanish to leading order.  We find, after a mess of algebra, that
\begin{eqnarray}
A_k^0 &=& \frac{N^2}{2k(N-k)} \nonumber\\
A_k^1 &=& -\frac{N(N-2k)}{2k(N-k)} \nonumber\\
A_k^2 &=& -\frac{N(k-1)}{2k} \\
B_k^0 &=& \frac{N}{2(N-k)} \nonumber\\
B_k^1 &=& -\frac{N(N-k-2)}{4(N-k)} \nonumber\\
C_k   &=& \frac{N-k-1}{N-k} \nonumber
\end{eqnarray}
We see from this that the profile is asymmetric, with the left (trailing)
side larger.  However this asymmetry is proportional to $1/x$ and so
slowly decays in time.

Now we plug all this into the equation of motion for $f$, finding that
\begin{equation}
\label{cofm}
\dot f = - v f' + D (x f)''
\end{equation}
where the velocity, $v$, is given by
\begin{equation}
\label{vel0}
v = \mu \frac{N(N-1)}{2(N+1)}
\end{equation}
and so increases (roughly) linearly with $N$, in accord with our general
expectations above.  The diffusion constant of
the center of mass, $D=\frac{N}{2(N+1)}$ is roughly constant with $N$.
The (unnormalized) long-time similarity solution for $f$ is
\begin{equation}
f(x,t) = \frac{1}{x}{\left(\frac{x}{t}\right)}^{v/D}e^{-x/Dt}
\end{equation}
We see that the spreading of the center of mass is linear in time.

Unfortunately, we see that the $g$'s are actually of order $O(\mu N)$,
so the expansion is not uniformly valid in $N$. Nevertheless, it is clear
that the {\em form} of the equation of motion is true to all orders in $\mu$, 
with
just the coefficients $v$ and $D$ varying.  The task therefore is to compute 
these
coefficients for arbitrary $\mu$.  For the case $N=2$, we can generalize 
our analysis above to accomplish this, so we return to that case.

\subsection{Two Individuals, Arbitrary $\mu$}
For the $N=2$ case, we can straightforwardly generalize our small $\mu$
solution given above to the case of arbitrary $\mu$. 
The relevant variables are $f_n(x,t)$, ($n=0,1,\ldots$),
the probability of an individual at $x$ and one at $x+n$. The
equations of motion are:
\begin{eqnarray}
\dot f_0(x) &=& \frac{1}{3}\left[-4\mu x f_0(x) 
+ x(1-\mu)\sum_{n=1}^\infty \left[f_n(x) + f_n(x-n)\right]+ \frac{1}{2}\mu\Big((x+1)f_1(x+1) + (x-1)f_1(x-1)\Big)\right]\nonumber\\
\dot f_1(x) &=& \frac{1}{3}\bigg[2\mu \Big(x f_0(x) + (x+1)f_0(x+1)\Big)  
- (1+\mu)(2x+1) f_1(x)  \nonumber \\
&\ &\ \ \ \ \ \ \ + \frac{1}{2}\mu\sum_{n=1}^\infty \Big[xf_n(x) + (x+1)f_n(x+1) + xf_n(x-n)
+(x+1)f_n(x-n+1)\Big]\nonumber\\
&\ &\ \ \ \ \ \ \ +  \frac{1}{2}\mu\Big((x+2)f_2(x) + (x-1)f_2(x-1)
\Big)\bigg]\nonumber\\
\dot f_n(x) &=& \frac{1}{3}\left[-(1+\mu)(2x+n) f_n(x) 
+ \frac{1}{2}\mu\Big((x+n+1)f_{n+1}(x) + (x-1)f_{n+1}(x-1)\Big)\right.\nonumber\\
&\ &\ \ \ \ + \left. \frac{1}{2}\mu\Big((x+n-1)f_{n-1}(x)+(x+1)f_{n-1}(x-1)
\Big)\right]\nonumber\\
\end{eqnarray}
where in the last equation, $n \ge 2$.
Again, there is an exact steady-state solution, with all the $f_n$ independent
of $x$.  If we choose the ansatz $f_n=A r^n f_0$, the equation for $\dot f_n$
reads
\begin{equation}
0=-\frac{1}{6}(2x+n)A f_0 r^{n-1}\Big(2(1+\mu)r + \mu r^2 + \mu\Big)
\end{equation}
which has the solution, (defining $\chi\equiv 1/\mu$)
\begin{equation}
r=\chi + 1 - \sqrt{\chi^2 + 2\chi}
\end{equation}
For small $\mu$, $r \approx \mu/2 \ll 1$, and $r$ increases monotonically with
$\mu$.  This is in accord with our expectation for a rapid falloff of $f_n$
with $n$ for small $\mu$. For the maximum physical 
$\mu=1$, $r = 2 - \sqrt{3} \approx 0.27$. 
Plugging this form for $f_n$ into either of the two remaining equations
yields $A=4(1-r)/(1-3r)$. The normalization condition that $\sum_n f_n=1$
then determines $f_0 =(1-3r)/(1+r)$.  Note that $f_0$ and $A$ are positive, 
as they should
be, for all physical values of $\mu$. We can then calculate the exact
squared-width of the bound state,
\begin{equation}
\label{omega}
<n^2>=\sum_n n^2 f_n = 2\mu \ .
\end{equation}
Note that the first-order result derived above in the small $\mu$ limit
is in fact exact to all orders!

We can again generalize to the long-time limit where the derivatives of
$f$ are small, but non-zero. We write
\begin{equation}
f_n = r^n (Af_0 + B_n f_0' + C_n f_0'/x + D_n f_0'')
\end{equation}
If we assume the expected form for $\dot f_0$:
\begin{equation}
\dot f_0 =  -v f_0' + D(xf_0)'' ;
\end{equation}
we can obtain a set of equations for the coefficients $B_n$, $C_n$ and $D_n$
along with $\tilde v$ and $D$
by matching the coefficients of $f_0$ and its derivatives.  After much
algebra, we find
\begin{eqnarray}
B_n&=& nA/2 \nonumber \\
C_n&=& \left(-\frac{1}{8}n^2 + \frac{(3\chi-1)r}{4(1-r^2)}n + 
\frac{r^2(2\chi-3)}{2(1-r)^2(1+r)(1-3r)}\right)A \nonumber \\
D_n&=& \left(\frac{1}{8}n^2 + \frac{(\chi-3)r}{4(1-r^2)}n +
\frac{r^2(2\chi-1)}{2(1-r)^2(1+r)(1-3r)}\right)A \\
v&=&\mu/3 \nonumber \\
D&=&\mu/3 \nonumber
\end{eqnarray}
Just as was the case with the width, the coefficient of the drift 
$\tilde v$ and diffusion $D$ are given exactly by the 
linear in $\mu$ expressions found above!
Such simple results cry out for a simpler derivation, which we now present.

\section{Moment Equations}
As we have seen, the dynamics of our discrete evolution model 
gives rise to a pulse of a typical width; the center of this pulse
propagates at long times with some velocity and also diffuses. 
Unfortunately, the methodology
used to obtain these results (essentially studying the full master equation)
cannot be used to obtain the macroscopic parameters of the velocity and
diffusion constant in general, and also do not indicate anything about
the pre-asymptotic state. What we will see now is that one can re-derive
the same picture for the long-time limit of the model at arbitrary parameters.
This derivation proceeds by working with moment equations for the probabilistic
process. These moment equations are valid at all times, and so also shed 
light on the short-time dynamics of the system.

\subsection{The Width of the Pulse}
Staying for the moment with the $N=2$ case,
let us focus on the average squared-width, $<n^2>$.
We wish to calculate $\dot{<n^2>}$, the rate of change of the squared-width
with time.
Given the definition of $<n^2>$ in terms of the $f_n$'s, (see Equation
(\ref{omega})) we can obtain what
we want from the equations of motion of the $f_n$'s.  We find
\begin{eqnarray}
\dot{<n^2>} &=& -\frac{2}{3}\sum_n {\bar x} n^2 f_n(x) + 
\frac{4}{3}\mu \sum_n {\bar x} f_n(x) \nonumber\\
 &=& -\frac{2}{3} <{\bar x}n^2> + \frac{4}{3}\mu <\bar x>
\end{eqnarray}
where $\bar x \equiv x+n/2$ is the average position of the two individuals
making up the configuration $f_n(x)$. If we may factorize the
expectation value $<{\bar x} n^2> \approx <{\bar x}><n^2>$, (a step we can
justify for long times from our explicit solution above) we get
\begin{equation}
\label{width2}
\dot{<n^2>} = -\frac{2}{3} <{\bar x}><n^2> + \frac{4}{3}\mu <\bar x>
\end{equation}
which immediately yields the steady-state value of $\omega = 2\mu$, which we
obtained in the previous section.  

This argument can be easily extended to arbitrary N.  We now define the
squared-width, $\omega$, as the mean squared distance between {\em pairs}
of individuals, taken over all $N(N-1)/2$ pairs. 
\begin{equation}
\omega\equiv \frac {2}{N(N-1)} \sum_{i>j} (x_i - x_j)^2
\end{equation}
and so its ensemble average is
\begin{equation}
<\omega>= \frac {2}{N(N-1)}\sum_C P_C \sum_{i>j} (x_i - x_j)^2
\end{equation}
where the first sum is over all configurations $C$, weighted by their
probability $P_C$. We are thus considering a 2-point function, 
instead of the full N-individual joint probability distribution.  
We can calculate $\dot{\omega}$ by considering the effect of each possible
event on $\omega$. We find, summing over
all events where $i$ is born and $j$ dies,
\begin{eqnarray}
\dot{<\omega>}= \frac{2}{N(N-1)(N+1)}\sum_C P_C \sum_{i,j} x_i &\ & 
\left[(1-\mu)\left(
\sum_{k\ne i,j}\left[(x_k - x_i)^2 - (x_k-x_j)^2\right] + 0 - (x_i - x_j)^2
\right)\right. \nonumber\\
&\ & \ \ \mbox{} + \frac{\mu}{2}\left(\sum_{k\ne i,j}\left[(x_k-x_i-1)^2 - (x_k-x_j)^2 \right] + 1 - (x_i-x_j)^2\right) \nonumber \\
&\ & \ \  \left.\mbox{}+ \frac{\mu}{2}\left(\sum_{k\ne i,j}\left[(x_k-x_i+1)^2 - (x_k-x_j)^2 \right] + 1 - (x_i-x_j)^2\right)\right] \nonumber \\
\end{eqnarray}
After some algebra, we get 
\begin{equation}
\label{widtheom}
\dot{<\omega>} = \frac{2N}{N+1} \left(- \frac{<{\bar x} \omega>}{N} + 
 <C_3> + \mu <\bar x> \right)
\end{equation}
where $\bar x$ is the average $x$ value of the population, $(1/N)\sum_i x_i$
 and 
$C_3$ is the asymmetry, $(1/N)\sum_i(x_i - \bar x)^3$.  
This asymmetry automatically
vanishes in the case of 2 particles, so the equation reduces to the
Equation (\ref{width2}) in that case.
We have already seen that the average asymmetry vanishes at long times like 
$1/t$, (an argument which easily extends to all orders in $\mu$), so that 
upon factorizing the expectation value $<{\bar x}\omega>$ as above, we get
that the width asymptotically approaches $\mu N$.

We can understand simply the origin of each of the three terms in the
equation of motion for $\dot{<\omega>}$.  The $\mu$ term
comes from the fact that mutation acts to increase the width.  Since only
babies can mutate, the effective mutation rate is proportional to
$\mu x$. The asymmetry term expresses the effect of selective pressure.
If there are particles that lag behind, giving rise to a negative
asymmetry, they are quickly killed off, decreasing the width. The last
term, $<\bar{x} \omega>$, is more subtle.  It expresses the fact that the 
population tends
to collapse even in the absence of selective pressure.  Death always
acts to reduce population variability, whereas birth does not increase
it.  This effect is crucial for the stabilization of the width against
the effects of mutation at long times.  The rate associated with
this effect is $2\bar x/(N+1)$.  As the velocity 
scales linearly in $N$, however, (which we have proved to lowest order in 
$\mu$ above and which we demonstrate below to all orders)
the time scale for the asymptotic width dynamics is essentially 
{\em independent} of $N$.  Unfortunately, this does not mean that
the asymptotic state is necessarily approached in a time which is $N$
independent; we will see later that this issue is actually considerably
more complex.

\subsection{The Drift Velocity}
It turns out that not only can we derive the width to all orders in $\mu$,
we can also simply relate this width to the velocity of the average mean 
position, $\dot{\bar x}$.

The average mean position is given by
\begin{equation}
<\bar x> = \frac{1}{N}\sum_{C} P_C\,\sum_i x_i
\end{equation}
Now a little reflection suffices to
convince oneself that mutation does not contribute to the instantaneous 
velocity, since the probabilities for mutating left and right are identical.
Thus the entire velocity is due to the collapse events. Then, as above, 
summing over all
events where $i$ has a baby, resulting in the death of $j$,
\begin{eqnarray}
\label{veltrue}
\dot{<\bar x>}&=& \frac{1}{N(N+1)}\sum_C \sum_i \sum_{j} P_C\, x_i(x_i - 
 x_j) \nonumber\\
&=& \frac{1}{N(N+1)}\sum_C \sum_i \sum_{j} P_C \,(x_i^2 - x_ix_j) \nonumber\\
&=& \frac{N-1}{2(N+1)} <\omega> 
\end{eqnarray}
Thus, the velocity is just exact proportional to $<\omega>$, the average
squared-width!
This is physically reasonable: the velocity arises from the collapse of the
population onto the more fit members.  The wider the distribution, the faster
the average fitness increases.  Now, we found
above the asymptotic value of $<\omega>$ to be ${\mu N}$, 
yielding the striking result
that the velocity is exactly linear in $\mu$, and given by our lowest
order result, Eq. (\ref{vel0}) found above!

Turning back to the simulation results in Fig. 2, we see that the results are
indeed consistent with a velocity converging to the theoretical value,
Eq. (\ref{vel0}).  Extending the run to longer times confirms this
result.  Similarly, we see the squared-width tracks with the velocity
as predicted, increasing toward the expected value $\mu N$. 

We can also calculate in a similar manner the diffusion constant $D$ of the 
center of 
mass motion.  From Eq. (\ref{cofm}) above, one can deduce that 
$d/dt \left(< ({\bar x})^2 > - < \bar x >^2 \right)= 2D <\bar x>$.  
Calculating this
quantity directly via the moments, we find
\begin{equation}
\frac{d}{dt}\left(< ({\bar x})^2 > - < \bar x >^2\right) = \frac{1}{N+1} 
 \left(<C_3> + \frac{N-1}{N}<\bar x \omega> + \mu <\bar x >\right) 
\end{equation}
Since, at long times, $<C_3>$ decays to zero, and $<\omega>$ approaches
$\mu N$, we get that 
\begin{equation}
D=\frac{\mu N}{2(N+1)}
\end{equation}
again reproducing our lowest order result exactly.

So, we have been able to show that the asymptotic state of the evolution process
is one in which the width saturates to a fixed value, leading to a pulse
propagating at constant velocity in fitness space. The exact values of the
velocity and width have been determined. What we still need to discuss is
the failure of mean-field theory and the related issue of the transient
behavior before the asymptotic state sets in. We will return to these issues
shortly. Now, we briefly detour to comment on the flat landscape case.

\section{The Flat Landscape and other Variants}
All the analysis above works as well (and more simply) in the case of
the flat landscape; i.e., where the birth rate is {\em independent} of
$x$. A variant of this flat-landscape model in which not only babies 
mutate, but every individual does, was considered first by Zhang, et. al
\cite{zhang} and exactly solved by Meyer, Havlin, and
Bunde\cite{Meyer}. In this variant, the relevant parameter is the ratio 
of the mutation rate to the birth rate, which we may take to be unity.
This ratio we label $\tilde \mu$, 
to distinguish
it from $\mu$, the {\em probability} that a baby mutates.  The context in which
this variant was analyzed was a problem of population migration, so that
the ``mutation'' occurs in physical space, not ``fitness'' space. One can
consider also an intermediate case, where everyone mutates (or migrates)
and the birth rate depends on position, (the ``emigration'' problem, where
people look for ``greener'' pastures).  All these variants have many features
in common.  As we shall see, the major differences between the models are 
1)the nature of the ``drift'', or velocity; and 2)the time-scale for the 
wave-function to reach its asymptotic width.

The flat landscape model of Zhang, et. al. and Meyer, et. al. is 
sufficiently simple to 
allow for a complete analytic solution, 
which Meyer et. al. present in a lovely
piece of work \cite{footnote}.  One essential difference in the
flat landscape case is that there is no drift.  The equation of motion for 
the center of mass
in the long-time limit is exactly the diffusion equation:
\begin{equation}
\dot f = D f''
\end{equation}
where the diffusion constant $D = \tilde \mu/2$ is exactly 
that of a single individual.  
The explanation for this fact, already noted by Zhang, et. al.,
is that the collapse event leaves the parent
unaffected, so that there is at long times one ``ur-individual'' who just
propagates as if nothing ever happened.  The localization of the wavefunction
then implies that all the diffusion of the center of the mass is due just
to the diffusion of this ``ur-individual''. Comparing this to our result we see that this parallels our results for
the diffusion constant in the ``tilted'' case.  There we found that the
diffusion constant was $N\mu x/2(N+1)$.  This is just the effective diffusion
constant for a single individual, since in the babies-only model, a diffusion
step only occurs if the baby is not immediately killed, so that the
effective rate of mutation is $\mu N/(N+1)$ times the birth rate $x$, giving rise
to a diffusion constant as noted above. The fact that only babies mutate
is not an essential change since the dynamics would be the same if only
the parent mutated, so again only the motion of the ``ur-individual'' need
be considered. 

\vspace{-1.5in}
\begin{figure}
\centerline{\epsfxsize = 6.in \epsffile{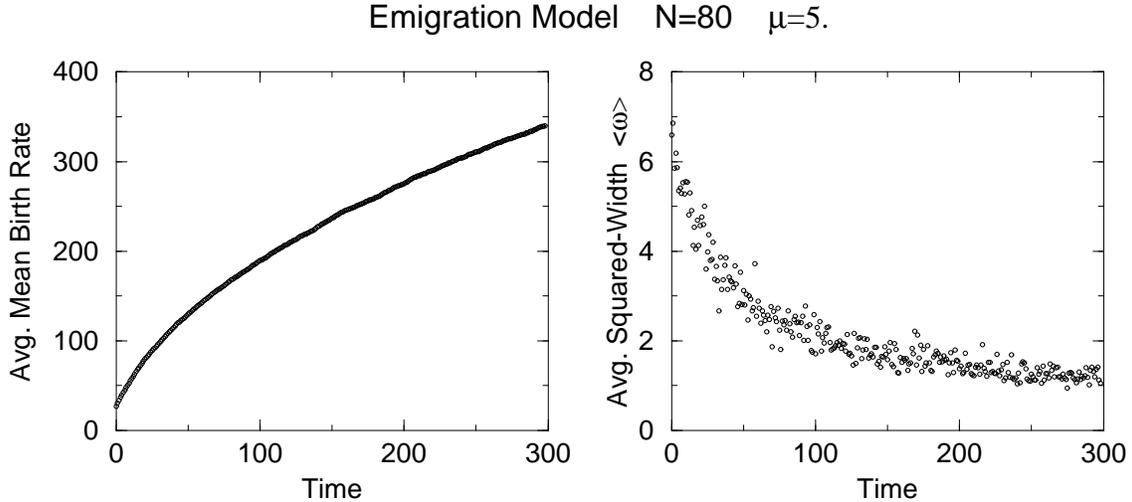}}
\caption{a) Mean birth-rate for population of $N=80$ individuals 
as a function of time in the ``emigration'' model, with mutation rate 
$\mu=5.$, averaged over 100 realizations. \ \ \ 
b) Width-squared from same simulation.}
\end{figure}

Of course, the drift must vanish in the flat-landscape case, by virtue of 
the left-right symmetry.
In the intermediate model where all individuals can mutate, but the landscape
is tilted, it is straightforward to see that the velocity is proportional
to $1/\bar x$, the current average position, so that the position should
scale as $\sqrt{t}$. This is related to the fact that the width of the wave 
function decreases as $1/{\bar x}$. The point is that a constant mutation
rate is not effective enough to combat the ever-increasing suppression of
variability due to the ever-increasing birth/death rate.  This behavior is
apparent in the simulation data of this model for $N=80$
presented in Fig. 3. We will return to a discussion of this ``emigration''
model in Sec. VII.

As already mentioned, the third order moment $C_3$ in the width equation
represents the effect of selective pressure. Is is therefore absent in
the flat landscape case; instead, the moment equation
for the width squared, $\omega$, reads
\begin{equation}
\dot{<\omega>} = 2 \left(- \frac{<\omega>}{N} + \tilde \mu \right)
\end{equation}
Also missing, naturally,  are the factors of $\bar x$.  Thus the moment
equation closes, allowing for an exact solution, identical to that
found by Meyer, et. al.

The last major  difference worth noting is the time scale for 
the asymptotic width dynamics.
We saw how in the tilted case the time scale is $\bar x/N$. The factor
$\bar x$ is the effective birth rate.  In the flat landscape case, the
birth rate is constant, so the time scale is fixed and or order $1/N$. Thus
stabilization of the width in the flat landscape case always takes place 
after $O(N)$ time.

\section{Moment Equations and Mean-Field Theory}

The equations of motion we have derived for the width and velocity not
only allow us to calculate the asymptotic behavior of these quantities.
They also allow us to make contact with the mean-field theory we presented
in section IIB.  The point is that one can derive parallel equations for these
moments directly from the mean-field (MF) theory. By comparing the exact
equations with the approximate ones obtained from the mean-field theory, we
can obtain some insight into the breakdown of the mean-field theory, and
how to improve upon it.

The MF equation for the velocity is easy to obtain.  The mean of $x$ is
\begin{equation}
<x>_{\text{MF}} = \sum n P(n)
\end{equation}
Using the MF equation of motion for $P$, Eq. (\ref{mfdisc}), we obtain
\begin{equation}
{\dot{<x>}}_{\text MF} = C_2^{\text{MF}}
\end{equation}
where $C_2^{\text MF}$ is the second cumulant, $<n^2>_{\text MF} - <n>_{\text MF}^2$.  This second cumulant is, to leading order in $1/N$, just $\omega/2$.
Thus, comparing with the exact equation, Eq. (\ref{veltrue}), we see that
the MF result is exact up to trivial $1/N$ corrections.

We can obtain in a similar manner an equation of motion for the second
cumulant. We find
\begin{equation}
{\dot C_2}^{\text MF} = C_3^{\text MF} + \mu <x>_{\text MF}
\end{equation}
Comparing with the exact equation for $\omega$, Eq. (\ref{widtheom})
we see the equations correspond except for the absence of the
stabilizing term $-<x C_2>/N$.  Now of course this term is formally
of order $1/N$, and so the MF cannot be expected to reproduce it \cite{doug}.
Nevertheless, we have seen that this is the term responsible for the
saturation of the width at long times. The point is that $<C_2>$ is $O(N)$
at long times and so the term is in truth not negligible.

The question at this point is, how long is long?  When does this and other
formally small terms become relevant?  The answer is very quickly, on a time
scale of $O(1)$.  The reason for this is the blowup of MF we alluded to
in section 2.  We can see this blowup directly by solving exactly the MF
equation of motion, Eq. (\ref{mfdisc}).  To do this, we define a generating
function for the moments, $Z(a,t)$,
\begin{equation}
Z(a,t)\equiv \sum_n P(n,t) e^{an}
\end{equation}
It is actually more convenient to work in terms of the function $F\equiv \ln Z$,
which is a generating function for the cumulants.  
Plugging the definition of $F$ in the equation of motion for $P$, we find
\begin{equation}
\dot F(a,t)= (1+\mu \cosh(a) - \mu)\frac{d}{da}F(a,t) - \frac{dF}{da}\bigg|_{a=0}
\end{equation}
We can solve this equation by means of defining a new variable $s$ such
that 
\begin{equation}
\frac{ds}{da}=\frac{1}{1 + \mu \cosh(a) - \mu}
\end{equation}
so that
\begin{equation}
s(a)=\frac{1}{\sqrt{1-2\mu}} \ln\left(\frac{1+\sqrt{1-2\mu}\tanh(a/2)}
{1-\sqrt{1-2\mu}\tanh(a/2)}\right)
\end{equation}
for $\mu < 1/2$ and
\begin{equation}
s(a)=\frac{2}{\sqrt{2\mu-1}} \arctan\left(\sqrt{2\mu-1}\tanh(a/2)\right)
\end{equation}
for $\mu > 1/2$.  In terms of s, the equation for $F$ reads
\begin{equation}
\dot F(s,t) = \frac{d}{ds} F(s,t) - \frac{dF}{ds}\bigg|_{s=0}
\end{equation}
whose solution is
\begin{equation}
F(s,t)=F_0(s+t) - F_0(t)
\end{equation}
where $F_0(s) \equiv F_0(a(s))$ is the generating function for the
cumulants of $P(n,t=0)$.  For example, if we start with a zero width
pulse located at $n_0$ at time $t=0$, $F_0(s)=n_0 a(s)$.  Then, translating
back to our original variable $a$, we find (for our zero width pulse)
\begin{equation}
F(a,t)=n_0 (s^{-1}(s(a) + t) - s^{-1}(t))
\end{equation}
The key feature of this solution is that $s_\infty\equiv s(a=\infty)$ 
is finite, so that the inverse function $s^{-1}$ is singular at $s_\infty$.
This causes $F$ to diverge at time $t^*=s_\infty$. Explicitly, 
\begin{equation}
t^*=\left\{ \begin{array}{ll}
\frac{1}{\sqrt{1-2\mu}}\ln(\frac{1+\sqrt{1-2\mu}}{1-\sqrt{1-2\mu}})\ \ \ \ \  
&\mu < \frac{1}{2} \\
\frac{2}{\sqrt{2\mu-1}}\arctan(\sqrt{2\mu-1})\ \ \ \ \  &\mu > \frac{1}{2} 
\end{array} \right. 
\end{equation}
so that $t^*$ diverges as $-\ln \mu$ for $\mu \ll 1$ and decreases 
monotonically to $\pi/2$ for the maximal $\mu = 1$.  For example, the
first moment of $P$, the mean position, is given by
\begin{equation}
<x>_{\text MF}=\frac{dF}{da}\bigg|_{a=0}=
\left\{ \begin{array}{ll}
\frac{n_0}{(1 - \frac{1}{\chi^2}\tanh^2(\frac{\chi t}{2}))\cosh^2(\frac{\chi t}{2})}\ \ \ \ \  
&\mu < \frac{1}{2} \\
\frac{n_0}{(1 - \frac{1}{\chi^2}\tan^2(\frac{\chi t}{2}))\cos^2(\frac{\chi t}{2})}\ \ \ \ \  
&\mu > \frac{1}{2} 
\end{array} \right. 
\end{equation}
where $\chi \equiv \sqrt{|1-2\mu|}$.  The divergence of this quantity at 
$t=t^*$ is apparent.

In fact, all nonzero cumulants
of $P$ diverge at $t^*$, with the $j$th cumulant diverging as $(t^*-t)^{-j}$.
This makes an analysis of the precise manner of the breakdown of MF rather
subtle.  Consider the equation of motion for $\omega$, Eq. (\ref{widtheom}).
It is true that $<\bar x>\omega/N$ is diverging as $t \to t^*$ as 
$(t^*-t)^{-3}$,
but then again the $C_3$ term in the equation is also diverging in exactly
the same manner.  As it is clear that MF must break down before $t=t^*$, it 
must be higher-order moments that are responsible.  These higher-order moments
are responsible for driving the low-order moments to diverge, and if
they are cut off somehow by the $1/N$ terms, the effect will eventually 
propagate down to the low-order moments.  

It is difficult at this stage
to pinpoint precisely the workings of this mechanism however.  At issue is 
whether there exists an intermediate time regime at very large $N$ where 
neither MF nor the asymptotic regime with its constant
velocity are accurate descriptions of the dynamics. To indicate the
nature of the problem, we present in Fig. 4 a graph of the time-development
of the squared-width, $<\omega>$, as a function of time as measured from 
simulations for a 
number of different $N$'s.  The initial time regime with its MF-like 
behavior is clearly evidenced for short times.  This rapid rise 
in the velocity is then
abruptly cut off.  The velocity then slowly increases to its asymptotic value.
The time-scale of this slow rise, and its dependence on $N$ are not at this
point clear to us.  We hope to return to these issues in a future work.

\section{The ``Emigration'' Model}
We now examine the breakdown of MF for the variant model where everyone
mutates, the ``emigration'' model.  As we indicated in Section VI, 
the asymptotic state is
characterized by a velocity decreasing as $1/\sqrt{t}$. The short-time
dynamics is controlled by the following MF equation:
\begin{equation}
\dot P(n) = (n-\bar n) P(n) + \frac{1}{2}\tilde \mu\left[P(n+1) - 2P(n) + P(n-1)\right]
\end{equation}
We can also solve this MF exactly.  The equation for $F(a)$ now reads
\begin{equation}
\dot F(a) = \frac{d}{da} F(a) + \tilde \mu (\cosh(a)-1) - \frac{dF}{da}\bigg|_{a=0}
\end{equation}
For an initial population located at $n_0$, so that $F(a,t=0)=n_0 a$,
one can verify that 
\begin{equation}
F(a,t)=\tilde \mu \Big(\sinh(a + t) -\sinh(t) - \sinh(a)\Big) + n_0 a
\end{equation}
satisfies the MF equation.  From this solution, one can read off
that the velocity $v=\tilde \mu \sinh(t)$ grows exponentially in time.
Thus, for this model, MF must break down by the time $ln(N/{\tilde \mu})$
since at this time the MF value of $<x><\omega>/N$ is competitive with
the $C_3$ term as given by MF. However, as with the only-babies-mutate model
analyzed in the previous section, the breakdown of MF appears to be more
complex.  Simulations indicate that in fact MF breaks down before this
time and in particular when $<x> <\omega>/N$ is still small compared
to the other terms.  We present in Fig. 5, simulations of the width as a function of
time for various $N$.  Again, for very short times, we see a regime where
MF is accurate.  The growth of the width is then suddenly cut-off and the
width starts decaying.  The cutoff is seen to take place at times earlier
than that indicated by the simple argument above, due to the role of 
higher-order moments, similar to what we found for the only-babies-mutate
model as discussed above.

\begin{figure}
\centerline{\epsfxsize = 3.in \epsffile{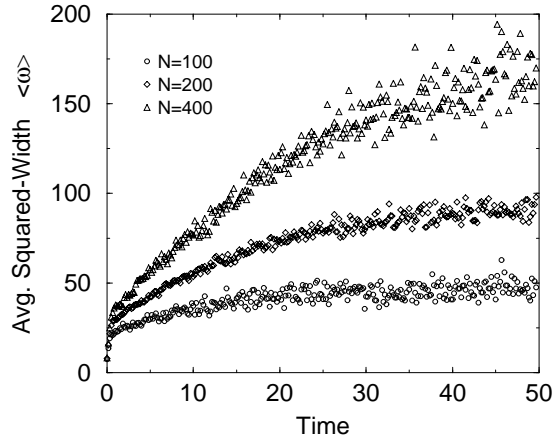}}
\caption{Squared-width as a function of time for populations of 
$N=100$ (circle), $N=200$ (diamond), and $N=400$ (triangle) individuals, 
with mutation rate 
$\mu=.5$, averaged over 160, 640 and 160 realizations respectively.}
\end{figure}

In our previous work\cite{virus1}, we presented
an analysis of a cut-off version of 
the MF for this everyone-mutates model.  The cut-off amounted to killing
the $n-\bar n$ term for those sites where $P(n)$ was less than some threshold
value\cite{cutoff}.  
This model possessed an asymptotic state which propagated at 
constant velocity.  In fact, starting from a distribution localized on
a single site $n_0$, the velocity in the cut-off theory increases
monotonically to this
asymptotic value.  We see that this cutoff theory seriously misrepresents
the actual physics, even at large $N$.  In essence, this form of cutoff
is too weak. We have seen that the corrections to MF are in fact
stabilizing in nature, whereas the simple truncation of this cutoff just
makes the growth at the leading edge marginal. The question of whether
a more physically correct cutoff can be constructed is intimately related
to the question of the precise nature of the breakdown of naive MF and the
rapidity of approach to the asymptotic state; these
questions we have chosen to postpone to future work.

\section{Conclusions}
The purpose of this paper was to present a detailed analysis of a simple
model for evolution on a smooth landscape. This model was originally
motivated by experiments on RNA viruses which showed persistent increases
in fitness without the ``punctuation" that should be present if the
fitness landscape had significant roughness thereby giving rise to
local maxima. Our results indicate that the system does settle into
a asymptotic state characterized by a constant rate of fitness growth.  
This state
is dominated by a balance between mutation and the
the loss of variability within the population due to death.  This latter effect, 
although formally weak in a large population, is crucial in rescuing the 
population from the runaway increase in birth-rate seen in the mean-field
limit.  

Our model exhibits certain generic features which should be sufficiently
robust that it is reasonable to hope that they apply also to the biological
system which motivated these investigations.  The most striking is the role 
of the population size on the dynamics.  Since the $N$-independent mean-field
treatment breaks down so spectacularly after a very short time, we may expect
that the population size is a very relevant parameter.  The other prediction
of our model that bears checking in the experimental system is the behavior 
of the population variance.  We have seen that if the initial variance is 
large, there is a rapid initial collapse of variance.  The variance then
asymptotes to a fixed quantity and does not exhibit diffusional broadening
in time.  Also, the fact that the rate of improvement of fitness vanishes
for small $N$ is very relevant for modelling the process of ``genetic 
bottlenecking'' \cite{bottle} used to create the initial population in the experiment.

Our investigations have implications beyond the narrow confines of the model
analyzed herein.  The breakdown of mean-field theory exhibited by our model 
is shared by diffusion-limited aggregation.  There also it is clear that this
breakdown is the result of the neglect of correlations, in this case 
correlations between the walker density and the density of the aggregrate.
This correlation is most apparent for particles who manage to outrun the typical
extent of the aggregate.  Whereas the ensemble-averaged walker density is 
quite large at the position of this outlying particle, the actual walker 
density is very small.  The neglect of this correlation leads to a blowup
of the mean-field exactly parallel to the blowup exhibited by the mean-field
studied herein.  The failure of simple cutoffs to correctly capture the 
asymptotic dynamics leads to the concern that similar cutoffs employed in the
study of mean-field DLA might also misrepresent the dynamics of the 
system.  

Our methods might also prove useful in the study of the behavior
of genetic algorithms\cite{GA}, where presents a similar system of evolving
populations.  Questions such as the speed of approach to the optimal state,
the dependence on population size and the
behavior of the population variance lie at the heart of understanding the
working of these algorithms.

DAK thanks S. Havlin for sharing the results of his work prior to publication
and I. Kanter for useful conversations.
HL and DR acknowledge the support of the US NSF under grant DMR94-15460;
LT was supported in part by DOE DE-FG03-95ER14516.  DAK and HL acknowledge the
support of the US-Israel Binational Science Foundation.  DAK acknowledges
the support of the Minerva Foundation.


\references
\bibitem{virus1}L. Tsimring, H. Levine, and D. A. Kessler, \prl {\bf 76},
4440 (1996). 
\bibitem{Holland95}I.S.Novella, E.A.Duarte, S.F.Elena, A.Moya,
E.Domingo, and J.J.Holland, {\em Proc. Natl. Acad. Sci. 
USA,} {\bf 92}, 5841-5844 (1995).
\bibitem{Eigen}M.Eigen and C.Biebricher, in: {\em RNA Genetics}, eds.
E.Domingo, J.J.Holland, and P.Ahlquist (CPC Press, Boca Raton, FL, 1988).
\bibitem{Schuster} W. Fontana, W. Schnabl and P. Schuster, {\em Phys. Rev. A}
{\bf 40}, 3301 (1989). 
\bibitem{Kaufman}S.A.Kaufman, {\em The Origins of Order}, Oxford
Univ.Press, New York, Oxford, 1993.
\bibitem{dla} T.A. Witten and L.M.Sander, \prb, {\bf 27}, 5686 (1983).
\bibitem{tu}E.Brener, H.Levine, Y.Tu, \prl, {\bf 66}, 1978 (1991).
\bibitem{cutoff} A similar idea  was put forth by T.B. Kepler
and A.S.Perelson, {\em Proc. Natl. Acad. Sci. USA}, {\bf 92}(18), 8219
(1995).
\bibitem{zhang}Y.-C. Zhang, M. Serva, and M. Polikarpov, J. Stat. Phys. 
{\bf 58}, 849 (1990).
\bibitem{Meyer}M. Meyer, S. Havlin, and A. Bunde, preprint, 1996.
\bibitem{footnote} Meyer, et. al. only analyze the 
large $\tilde \mu$ limit, where the random walk of the individuals can be
approximated by a diffusion equation. However, it is straightforward to
generalize their analysis to the general $\tilde \mu$ case. The conclusions 
of such an analysis agree exactly with the results of the
analysis in the text.
\bibitem{doug} It is possible to show that this $1/N$ term does appear if
one derives an equation coupling the mean field to the two-particle
distribution function; this result will be presented elsewhere.
\bibitem{bottle}H. J. Muller, {\em Mut. Res.} {\bf 1}, 2 (1964).
\bibitem{GA}D. E. Goldberg, {\em Genetic Algorithms in Search, Optimization,
and Machine Learning}, Addison-Wesley, Reading, MA., 1989.
\end{document}